\documentclass{article}
\usepackage{epsfig}
\def\H{H\hskip-8.5pt/\hskip2pt}

\def\VEV#1{\left\langle #1\right\rangle}

\def\lsim{\mathrel{\mathpalette\@versim<}}

\def\gsim{\mathrel{\mathpalette\@versim>}}

\begin{document} 

\begin{titlepage} 

\begin {centering} 

{\Large {\bf CPT and Decoherence in Quantum Gravity} } \\

\vspace{2cm}

{\bf Nikolaos E. Mavromatos }\\
\vspace{0.5cm}
King's College London, Department of Physics, \\ London, Strand WC2R 2LS, U.K


\vspace{3cm} 

{\bf Abstract } 

\end{centering} 
\vspace{1cm} 
In this review, I first discuss briefly some theoretical motivations
for potential Lorentz Violation and deviation from ordinary quantum
mechanical behavior (decoherence) of field theoretic systems in the background of some quantum gravity (QG) models.
Both types of effects lead to CPT violation, but they can be disentangled experimentally.
I, then, proceed to a description of
precision tests of CPT symmetry using neutral and charged
Kaons, which are of direct relevance to the main theme of this conference.
I emphasize
the potentially unique r\^ole of neutral meson
factories in providing ``smoking-gun'' evidence of some
QG-decoherence models in which the CPT
quantum mechanical operator is not well defined. This is achieved by means of
potential observations of QG-induced modifications of the pertinent  Einstein-Podolsky-Rosen (EPR) particle correlations.

\vspace{2cm}

\begin{centering} 

{\it Kaon International Conference (Kaon07), 
         May 21-25, 2007, \\
         Laboratori Nazionali di Frascati dell'INFN, Italy.}

\end{centering} 

\end{titlepage} 

\section{Lorentz Violation and Decoherence from Quantum Gravity: Motivations}

Any quantum theory, formulated on flat space times, is symmetric
under the combined action of CPT transformations, provided the
theory respects (i) Locality, (ii) Unitarity (i.e. conservation of
probability) and (iii) Lorentz invariance. This is the celebrated CPT
theorem~\cite{cpt}. An extension of this theorem to Quantum Gravity (QG)
is by no means an obvious one; there may be information loss, in
certain space-time foam backgrounds~\cite{hawking}, implying an
evolution from pure to mixed quantum states, and hence
decoherence~\cite{hawking,ehns}. In such  situations {\it particle
phenomenology} has to be reformulated~\cite{ehns,poland} by viewing
our low-energy world as an open quantum system.
A similar situation might be encountered in Cosmologies with a cosmological constant, a model that seems to be favored by current astrophysical data on the acceleration of the Universe. Such models are characterized by cosmic (de Sitter) horizons, and again asymptotic states cannot be defined, and one may face a decoherence situation as a result of environmental degrees of freedom beyond the horizon (this issue however is still wide open, as the nature of the ``microstates'' of the de Sitter system is not understood at present).
In all such cases the \$ matrix, connecting
asymptotic \emph{in} and \emph{out} density matrices
rather than pure-state wave vectors,
$\rho_{\rm out} =\$ \rho_{\rm in} $
is {\it not invertible},
and this implies, by means of a theorem due to R. Wald~\cite{wald}, that the
CPT operator itself is \emph{not well defined}, at least from an
effective field theory point of view. This is a strong form of CPT
Violation (CPTV). This form of CPTV introduces a fundamental arrow of time/microscopic time
irreversibility, unrelated in principle to CP properties.
However, this arrow may not be observable experimentally,
if the experimentalist has access to the so-called decoherence-free subspaces,
which can be achieved, for instance, if the CPTV effects cancel out between
particle and antiparticle sectors. This
leads to a \emph{weak} form of CPT invariance~\cite{wald}.
This is a model dependent statement, and therefore subject to experimental
verification in principle.
Within the
scope of the present talk I will restrict myself to decoherence
and CPT invariance tests within neutral
Kaons~\cite{ehns,lopez,huet,benatti}. This type of (decoherence-induced) CPTV
exhibits some fairly unique effects in $\phi$ ($B$-meson, ...)
factories~\cite{bmp}, associated with a potential modification of
the Einstein-Podolsky-Rosen (EPR) correlations of the entangled
neutral Kaon ($B$-meson, ...) states produced after the decay of the
$\phi$-(or $\Upsilon$-, ...) meson.

Another fundamental reason for CPTV in QG
is the {\it spontaneous breaking of Lorentz symmetry
(SBL)}~\cite{sme}, without necessarily implying decoherence. In this
case the ground state of the field theoretic system is characterized
by non trivial vacuum expectation values of certain tensorial
quantities, $\langle {\cal A}_\mu \rangle \ne 0$,
or $\langle {\cal B}_{\mu_1\mu_2\dots}\rangle \ne
0~$.
A concrete example of SBL may be provided by string field theory models of open bosonic strings~\cite{kosstring}. In such models, there are cubic terms in an effective low-energy (target-space) Lagrangian involving the
tachyonic scalar field $T$, that characterizes the bosonic string vacuum, and invariant products of higher-tensor fields that appear in the mode expansion of a string field,
$T {\cal B}_{\mu_1 \dots \mu_n}{\cal B}^{\mu_1 \dots \mu_n}~. $
The negative mass squared tachyon field, then, acts as a Higgs field in
such theories, acquiring a vacuum expectation value, which, in turn, implies
non-zero vacuum expectation values for the tensor fields ${\cal B}$, leading in this way to energetically preferable configurations that are Lorentz Violating (LV). From the point of view of string theory landscape these are perfectly acceptable vacua~\cite{kosstring}.
An effective target-space field theory framework to discuss the phenomenology
of such LV theories is the so-called Standard Model Extension (SME). For our purposes in this work, the upshot of SME is that there is a \emph{Modified Dirac Equation}
for spinor fields $\psi$, representing leptons and quarks with
charge $q$:
{\small \begin{eqnarray}\left( i\gamma^\mu D^\mu - M -  a_\mu \gamma^\mu
- b_\mu \gamma_5
\gamma^\mu  -
\frac{1}{2}H_{\mu\nu}\sigma ^{\mu\nu} +
ic_{\mu\nu}\gamma^\mu D^\nu + id_{\mu\nu}\gamma_5\gamma^\mu D^\nu
\right)\psi =0~,
\end{eqnarray}}
where $D_\mu = \partial_\mu - A_\mu^a T^a - qA_\mu$ is an
appropriate gauge-covariant derivative. The non-conventional terms
proportional to the coefficients $a_\mu,~  b_\mu,~ c_{\mu\nu},~
d_{\mu\nu},~ H_{\mu\nu}, \dots $, stem from the corresponding local
operators of the effective Lagrangian which are phenomenological at
this stage. The set of terms pertaining to $a_\mu~, b_\mu$ entail
CPT and Lorentz Violation, while the terms proportional to
$c_{\mu\nu}~, d_{\mu\nu}~, H_{\mu\nu}$ exhibit Lorentz Violation
only. It should be stressed that, within the SME framework (as also with
the decoherence approach to QG), CPT violation does \emph{not
necessarily} imply  mass differences between particles and
antiparticles.

Some remarks are now in order, regarding the form and
order-of-magnitude estimates of the Lorentz and/or CPT violating
effects. In the approach of \cite{sme} the SME
coefficients have been taken to be constants. Unfortunately there is
not yet a detailed microscopic model available, which would allow
for concrete predictions of their order of magnitude.
Theoretically, the (dimensionful, with dimensions of energy) SME
parameters can be bounded by applying renormalization group and
naturalness assumptions to the effective local SME Hamiltonian,
which leads to bounds on $b_\mu$ of order $10^{-17}~{\rm
GeV}$. At present all SME parameters should be considered as
phenomenological and to be constrained by experiment.
In general, however, the constancy of the SME coefficients may not
be true. In fact, in certain string-inspired or stochastic models of
space-time foam that violate Lorentz symmetry~\cite{poland,bms}, the coefficients
$a_\mu, b_\mu ...$ are probe-energy ($E$) dependent, as a result of
back-reaction effects of matter onto the fluctuating space-time.
Specifically, in stochastic models of space-time foam, one may find~\cite{bms}
that on average there is no Lorentz and/or CPT violation, i.e., the
respective statistical v.e.v.s (over stochastic space-time
fluctuations)
 $\langle a_\mu~, b_\mu \rangle = 0~,$ but this is not true for
 higher order correlators of these quantities (fluctuations), i.e.,
$\langle a_\mu a_\nu \rangle \ne 0,~
 \langle b_\mu a_\nu \rangle \ne 0,~  \langle b_\mu b_\nu \rangle \ne 0~, \dots
 $. In such a case the SME effects will be
much more suppressed, since by dimensional arguments such
fluctuations are expected to be at most of order $E^4/M_P^2$, probably with
no chance of being observed in immediate-future facilities, and
certainly not in neutral kaon systems in the foreseeable future.

\begin{figure}[ht]
\centering
\includegraphics[width=6.5cm]{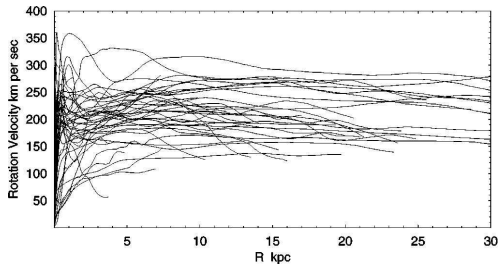}
\caption{Collage of Rotational Curves of nearby spiral galaxies obtained by combining Doppler data from CO molecular lines for the central regions, optical lines for the disks, and HI 21 cm line for the outer (gas) disks.
Graph from Y. Sophue and V. Rubin, Ann. Rev. Astron. Astrophys., Volume {\bf 31} (2001), 127.}
 \label{rc}
 \end{figure}
\begin{figure}[ht]
\begin{center}
\includegraphics[width=5cm]{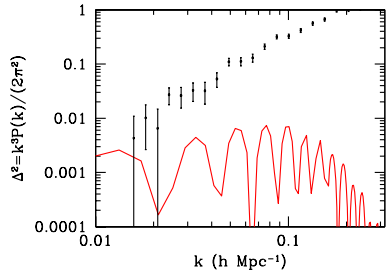}
\end{center}
\caption{Power spectrum $\Delta^2=k^3P(k)/(2\pi^2)$ vs. the scale $k$ of matter fluctuations (red curve, with wiggles) in a theory without dark matter as compared to observations of the galaxy
power spectrum. }
\label{baryongs}
\end{figure}

We mention at this stage that LV theories have been recently invoked in cosmology as a
way to
bypass the dark matter problem, by providing
relativistic field theories of gravity (TeVeS-like models)~\cite{teves}, with isotropic vector LV fields, $a_\mu (t)$,
with only $a_0 (t) \ne 0$,
which are such that: (i) at galactic scales, and for small gravitational accelerations, $g \le (200 {\rm km\,sec}^{-1})^2/(10\,{\rm kpc})$,
they result in Modified Newtonian Dynamics of the type proposed in \cite{milgrom},
leading to an experimental reproduction of the observed rotational curves of galaxies (see fig.~\ref{rc}), without the postulate of dark matter halos,
and (ii) their vector-field instabilities exhibit~\cite{liguori}  enhanced cosmic growth at galactic scales, in agreement with observations (see fig.~\ref{baryongs}), obviating once more the need for dark matter.

\begin{figure}[t]
\begin{centering}
  \epsfig{file=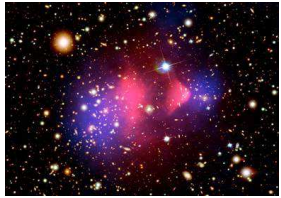, width=0.3\textwidth}
\caption{The Bullet cluster of Galaxies: the blue areas
indicate Dark Matter inferred by
Gravitational Lensing
Techniques, whilst the red areas denote luminous matter detected by x-rays.
This galaxy provides an example where the Modified Newtonian Dynamics Theory might be in trouble.} \label{bullet}
\end{centering}
\end{figure}

Although it is still unclear whether such models can fit all the available galactic and cosmological data,
in particular data from the bullet cluster of galaxies (see fig.~\ref{bullet})
and/or cosmic microwave background (CMB) data, nevertheless there are
recent claims that the CMB acoustic peaks could be fitted in the framework of such TeVeS-like models~\cite{mota}, provided that, in addition to the cosmological Lorentz Violation, hot dark matter
of massive neutrinos (of order 15\%) is present. Thus the issue of cosmological Lorentz Violation may still be considered as open, which could be resolved by particle physics tests. The latter can in principle provide upper bounds
for the LV effects, which then could be compared
with the ones required by observational cosmology, in order for LV to play an alternative r\^ole to Dark Matter. In this talk I  will restrict myself
to Lorentz symmetry tests using neutral Kaons~\cite{sme}, and discuss the most recent bounds.

I must stress at this point that QG-decoherence and Lorentz
Violation  are in principle independent~\cite{poland}. The
important difference of CPT violation in SBL models of Quantum
Gravity from that in space-time foam situations lies on the fact
that in the former case the CPT operator is well defined, but it
\emph{does not commute} with the effective Hamiltonian of the matter
system. In such cases one may parametrize the Lorentz and/or CPT
breaking terms by local field theory operators in the effective
lagrangian, leading to a construction known as the ``standard model
extension'' (SME)~\cite{sme}, which is a framework to study
precision tests of such effects. Frame dependence is important in disentangling LV effects from Lorentz invariant models.
In certain circumstances one may also violate locality,
but I will not discuss this case
explicitly here. Of course violations of locality could also be
tested with high precision by means of a study of discrete
symmetries in meson systems.
I must stress that the phenomenology of CPT violation is
complicated, and there seems \emph{not} to be a \emph{single} figure
of merit for it. Depending on the precise way by which
CPT violation is realized in a
given class of models of QG, there are different ways by which we
can test the violation~\cite{poland}. I stress again that within the above
frameworks, CPT violation does \emph{not necessarily} imply mass
differences between particles and antiparticles.

\section{Lorentz Violation and Neutral
Kaons}

\begin{figure}[htb]
\begin{centering}
  \epsfig{file=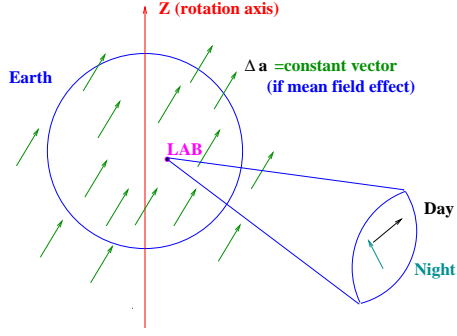, width=0.5\textwidth}
\caption{Schematic representation of searches for sidereal
variations of the CPT-violating parameter $\delta_K$ in the SME
framework. The green arrows, crossing the Earth indicate a constant
Lorentz-violating vector that characterizes the Lorentz-violating
ground state.} \label{ameas}
\end{centering}
\end{figure}

I commence the discussion
by a very brief description of experimental tests of Lorentz
symmetry, within the SME framework, using neutral Kaons, both
single~\cite{sme} and entangled states in a $\phi$
factory~\cite{adidomenico,testa}. In order to isolate
the terms in SME effective Hamiltonian that are
pertinent to neutral Kaon tests, one should notice~\cite{sme}
that
the relevant CPTV and LV parameter $\delta_K$ must be flavour diagonal,
C violating but P,T preserving, as
a consequence of strong interaction properties in neutral meson
evolution.
This implies that $\delta_K $ is
sensitive only to the $-a_\mu^q {\overline q} \gamma_\mu q$ quark
terms in SME~\cite{sme},
where $a_\mu$ is a Lorentz and CPT violating parameter,
with dimensions of energy, and $q$ denote quark fields, with the meson
composition being denoted by $M = q_1{\overline q}_2$.
The analysis
of \cite{sme}, then, leads to the following relation of the
Lorentz and CPT violating parameter $a_\mu$ to the CPT violating
parameter $\delta_K$ of the neutral Kaon system:
\begin{equation}
\delta_K \simeq i{\rm sin}\widehat{\phi} {\rm
exp}(i\widehat{\phi}) \gamma \left(\Delta a_0 - {\vec \beta_K}\cdot
\Delta {\vec a}\right)/\Delta m,
\end{equation}
with the short-hand
notation $S$=short-lived, $L$=long-lived,
$\Delta m = m_L - m_S$, $\Delta \Gamma = \Gamma_S - \Gamma_L$,
$\widehat \phi = {\rm arc}{\rm tan}(2\Delta m / \Delta\Gamma), \quad
\Delta a_\mu \equiv a_\mu^{q_2} - a_\mu^{q_1}$, and $\beta_K^\mu =
\gamma (1, {\vec \beta}_K)$ the 4-velocity of the boosted~ kaon, with
$\gamma$ the Lorentz factor.
The experimental bounds of $a_\mu$ in neutral-Kaon
experiments are based on searches of sidereal variations of
$\delta_K$ (day-night effects), as in fig.~\ref{ameas}.

From KTeV experiment~\cite{ktev} the following bounds of the $X$ and
$Y$ components of the $a_\mu$ parameter have been obtained
$\Delta a
_X, \Delta a _Y < 9.2 \times 10^{-22}~ {\rm GeV}$, where $X,Y,Z$
denote sidereal coordinates.
Complementary measurements for the $a_Z$ component can come from
$\phi$ factories~\cite{adidomenico}.

In the case of $\phi$-factories
there is additional dependence of the CPT-violating parameter
$\delta_K$ on the polar ($\theta$) and azimuthal ($\phi$) angles
{\small \begin{eqnarray} && \delta_{K}^\phi (|\vec p |, \theta, t) =
\frac{1}{\pi}\int_0^{2\pi}d\phi \delta_K(\vec p, t) \quad \simeq  \quad i{\rm sin}\widehat{\phi} {\rm exp}(i\widehat{\phi})
(\gamma/\Delta m) \cdot \nonumber
\\
&& \cdot \left(\Delta a_0 + \beta_K  \Delta {a_Z}{\rm
cos}\chi {\rm cos}\theta
+
\beta_K  \Delta {a_X}{\rm sin}\chi {\rm cos}\theta {\rm
cos}(\Omega t) + \nonumber \right.\\ 
&& \left. \beta_K  \Delta {a_Y}{\rm sin}\chi {\rm cos}\theta
{\rm sin}(\Omega t) \right)  \end{eqnarray}}
where $\Omega $
denotes  the Earth's sidereal frequency, and $\chi$ is the angle between
the laboratory Z-axis and the Earth's axis.
The experiment KLOE at DA$\Phi$NE is sensitive to $a_Z$: limits on
$\delta (\Delta a_Z)$ can be placed from forward-backward asymmetry
measurements $A_L = 2{\rm Re}\epsilon_K - 2 {\rm Re}\delta_K$. For
more details on the relevant experimental bounds we refer the reader
to the literature~\cite{adidomenico,testa}.
We only mention at this stage that in an upgraded DA$\Phi$NE
facility, namely experiment KLOE-2 at DA$\Phi$NE-2, the expected
sensitivity is~\cite{adidomenico} $\Delta a _\mu = {\cal
O}(10^{-18})$~GeV which, although not competitive with the
current KTeV limits on $a_{X,Y}$ given above, nevertheless constitutes an
independent complementary measurement of the $a_Z$ component.
Moreover, by looking at semileptonic decays, KLOE experiment can place limits
(of order $10^{-18}$) on the time components $\Delta a_0$ of this Lorentz and CPT Violating parameter~(A. DiDomenico, private communication),
which, as we discussed in the previous section might be the only non zero component in an isotropic cosmological model of Lorentz Violation.
Of course, it might well be that LV/CPTV effects cancel out between particle-antiparticle sectors, in which case they will be unobservable.
This is a model dependent statement. Finally, I mention that other precision tests can be performed using other meson factories (B-mesons, {\it
etc.}... ), which would also allow one to test the universality of QG
Lorentz-violating effects, if observed.

\section{Quantum Gravity Decoherence and  Neutral Kaons}

Quantum Gravity may induce decoherence and oscillations between Neutral-Kaon states $K^0 \leftrightarrow {\overline
K}^0$~\cite{ehns,lopez}, thereby implying a two-level quantum
mechanical system interacting with a QG ``environment''. Upon the
general assumptions of average energy conservation and monotonic
entropy increase, and the specific (to the Kaon
system) assumption about the respect of the
$\Delta S=\Delta Q$ rule by the QG medium,
the modified evolution equation for the reduced density matrices, $\rho$,
of the Neutral-Kaon matter reads~\cite{ehns}:
\begin{equation}
\partial_t \rho = i[\rho, H] + \delta\H \rho~, \quad {\delta\H}_{\alpha\beta} =\left( \begin{array}{cccc}
 0  &  0 & 0 & 0 \\
 0  &  0 & 0 & 0 \\
 0  &  0 & -2\alpha  & -2\beta \\
 0  &  0 & -2\beta & -2\gamma \end{array}\right)~.
\end{equation}
where $H$ denotes the hamiltonian of the Kaon system, that may contain
(possible) CPTV differences of masses and widths between particles and
antiparticles~\cite{lopez}, and $\delta\H $ is the decoherence matrix.
Positivity of $\rho$ requires: $\alpha, \gamma  > 0,\quad
\alpha\gamma>\beta^2$. Notice that $\alpha,\beta,\gamma$ violate
{\it both}  CPT, due to their decoherening nature~\cite{wald}, and
CP symmetry, as they do not commute with the CP operator
$\widehat{CP}$~\cite{lopez}: $\widehat{CP} = \sigma_3 \cos\theta +
\sigma_2 \sin\theta$,$~~~~~[\delta\H_{\alpha\beta}, \widehat{CP} ]
\ne 0$.
As pointed out in
\cite{benatti}, however,
in the case of $\phi$-factories complete positivity
is guaranteed within the above (single-particle) framework
only if the further conditions $\alpha = \gamma~ {\rm and} ~\beta = 0 $
are imposed.  Experimentally the complete positivity hypothesis,
and thus the above framework,
can be tested explicitly by keeping all three parameters.
In what follows, as far as single Kaon states are
concerned, we shall keep the $\alpha,\beta,\gamma$
parametrization~\cite{lopez},
and give the available experimental bounds for these parameters.
\begin{figure}[htb]
\centering
\epsfig{file=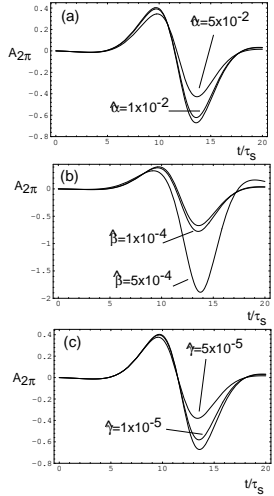, width=0.3\textwidth}
\vspace*{-0.5cm}\caption{Typical neutral kaon decay asymmetries
$A_{2\pi}$~\cite{lopez} indicating the effects of QG
induced decoherence.} \label{AT}
\end{figure}
The relevant observables are defined as $ \VEV{O_i}= {\rm Tr}\,[O_i\rho] $. One looks at the time evolution of decay asymmetries~\cite{lopez}
(see fig.~\ref{AT} for the case of $2\pi$ final states). The
important point to notice is that the two types of CPTV, within
and outside
quantum mechanics, can be
{\it disentangled experimentally}~\cite{lopez}.
We next mention that, typically, for instance when the final states
are $2\pi$, one has  a time evolution of the decay rate $R_{2\pi}$:
$ R_{2\pi}(t)=c_S\, e^{-\Gamma_S t}+c_L\, e^{-\Gamma_L t} + 2c_I\,
e^{-\Gamma t}\cos(\Delta mt-\phi)$, where $S$=short-lived,
$L$=long-lived, $I$=interference term, $\Delta m = m_L - m_S$,
$\Delta \Gamma = \Gamma_S - \Gamma_L$, $\Gamma =\frac{1}{2}(\Gamma_S
+ \Gamma_L)$. One may thus define the {\it Decoherence Parameter}
$\zeta=1-{c_I\over\sqrt{c_Sc_L}}$, as a (phenomenological) measure
of quantum decoherence induced in the system. In our decoherence
scenario, $\zeta$ corresponds to a particular combination of the
decoherence parameters~\cite{lopez} $ \zeta \to \frac{\widehat
\gamma}{2|\epsilon ^2|} - 2\frac{{\widehat \beta}}{|\epsilon|}{\rm
sin} \phi~,$ with $\widehat{\gamma} =\gamma/\Delta
\Gamma $, \emph{etc}.
The CPLEAR measurements gave the following bounds~\cite{cplear}
$\alpha < 4.0 \times 10^{-17} ~{\rm GeV}~, ~|\beta | < 2.3. \times
10^{-19} ~{\rm GeV}~, ~\gamma < 3.7 \times 10^{-21} ~{\rm GeV} $,
which are not much different from theoretically expected values in
some optimistic scenarios~\cite{lopez} $\alpha~,\beta~,\gamma =
O(\frac{E^2}{M_{P}})$. The experiment KLOE at Da$\Phi$NE updated
these limits recently by measuring for the first time the $\gamma$
parameter for entangled Kaon states~\cite{adidomenico,testa}:
$\gamma_{\rm KLOE} = (1.1^{+2.9}_{-2.4} \pm 0.4) \times
10^{-21}~{\rm GeV}$, as well as the (naive) decoherence parameter
$\zeta$. This bound can be improved by an order of magnitude in
upgraded facilities, such as KLOE-2~\cite{adidomenico}.

\section{Decoherence-CPTV and Modified EPR Correlations
of Entangled Neutral Kaons}

\begin{figure}[htb]
\centering
  \epsfig{file=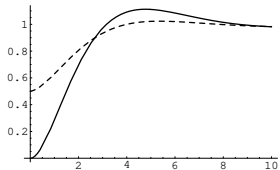, width=0.3\textwidth}
\caption{A characteristic case of the intensity
$I(\Delta t)$, with $|\omega|=0$ (solid line), vs $\Delta t$,
compared with (dashed line) $I(\Delta t)$, with $|\omega|=|\eta_{+-}|$,
$\Omega = \phi_{+-} -
0.16\pi$, for definiteness~\cite{bmp}.}
\label{intensomega}\end{figure}

If CPT is \emph{intrinsically} violated,
in the sense of being not well defined due to decoherence~\cite{wald},
the Neutral mesons $K^0$ and ${\overline K}^0$ should \emph{no
longer} be treated as \emph{identical particles}. As a
consequence~\cite{bmp}, the initial entangled state in $\phi$
factories $|i>$, after the $\phi$-meson decay, reads:
{\scriptsize \begin{eqnarray}
 |i> = {\cal N} \bigg[ \left(|K_S({\vec
k}),K_L(-{\vec k})>
- |K_L({\vec k}),K_S(-{\vec k})> \right)
 +  \omega \left(|K_S({\vec k}), K_S(-{\vec k})> - |K_L({\vec
k}),K_L(-{\vec k})> \right)  \bigg] \nonumber
\end{eqnarray}}
where $\omega = |\omega |e^{i\Omega}$ is a complex parameter,
parametrizing the intrinsic CPTV modifications of the EPR
correlations (``$\omega$-effect'').
The $\omega$-parameter controls the amount of contamination of the
final C(odd) state by the ``wrong'' (C(even)) symmetry state.
The appropriate observable (see fig.~\ref{intensomega})
is the ``intensity'' $I(\Delta t)
= \int_{\Delta t \equiv |t_1 - t-2|}^\infty
|A(X,Y)|^2$, with $A(X,Y)$ the appropriate $\phi$ decay
amplitude~\cite{bmp},
where one of the Kaon products decays to
the  final state $X$ at $t_1$ and the other to the final state $Y$
at time $t_2$ (with $t=0$ the moment of the $\phi$ decay).

The KLOE experiment at DA$\Phi$NE has released the first measurement of the
$\omega$ parameter~\cite{adidomenico,testa}: $ {\rm Re}(\omega) =
\left( 1.1^{+8.7}_{-5.3} \pm 0.9\right)\times 10^{-4}~$, ${\rm
Im}(\omega) = \left( 3.4^{+4.8}_{-5.0} \pm 0.6\right)\times
10^{-4}$. At least an order of magnitude improvement is expected
for upgraded facilities such as KLOE-2 at (the upgraded)
DA$\Phi$NE-2~\cite{adidomenico}. This sensitivity is not far from
certain optimistic models of space time foam leading to
$\omega$-like effects~\cite{bms}.
The $\omega$ effect can be
disentangled experimentally from \emph{both}, the C(even) background
- by means of different interference with the C(odd) resonant
contributions, and the decoherent evolution ($\alpha = \gamma$)
effects~\cite{bmp} - due to different structures.
Finally, I close this section by mentioning that, if this type of intrinsic CPT violation is due to a LV decoherent model, then
this should show up in a different size of the effect (if observed)
in B-factories, where the products of the decay of the $\Upsilon$-meson
are boosted as compared to those of the decay of the $\phi$-meson in $\phi$-factories, which occurs at rest. As far as $B$-factories are concerned, I also mention that the $\omega$-effect leads to intrinsic limitations for the accuracy of flavor tagging~\cite{tagging}.

\section{Precision T, CP and CPT Tests with Charged Kaons}

Precision tests of discrete symmetries can also be
performed with charged Kaons, as a result of the (recently
acquired) high statistics at the NA48 experiment~\cite{NA48}, in
certain decay channels, for instance $K^\pm \rightarrow \pi^+ + \pi^- + \ell ^\pm + \nu_\ell
(\overline \nu_\ell)$, abbreviated as $K_{\ell 4}^\pm$. One can
perform independent precision tests of T, CP and CPT using this
reaction~\cite{wu}, by comparing the decay rates of the $K^+$ mode with the
corresponding decays of the $K^-$ mode, as well as tests of $\Delta S =
\Delta Q$ and $|\Delta I|=1/2$ isospin rules. If CPT is violated,
through microscopic time irreversibility~\cite{wald}, then the phase
space analysis for the products of the reaction, from which one
obtains the di-pion strong-interaction phase shifts, needs to be
modified~\cite{wu}.
I would like to finish by mentioning the possibility of
exploiting the recently attained high statistics for charged Kaons
in the NA48 experiment~\cite{NA48} so as to use appropriate
combinations of \emph{both} reaction modes $K_{\ell 4}^\pm$ for
precision tests of physics beyond the Standard Model (SM), such as
supersymmetry, \emph{etc.}, including possible CPT violations. One
could look at T-odd triple momentum correlators~\cite{triple}:~ $\vec
p_\ell \cdot (\vec p_{\pi_1} \times \vec p_{\pi_2})$. The so
constructed CP-violating observables are independent of the lepton
polarization and thus easier to measure in a high statistics
environment, such as the NA48 experiment~\cite{NA48}.

\section*{Acknowledgements}

 I thank the organizers of the Kaon07
Conference, in particular A. DiDomenico, for the invitation and hospitality during this very interesting
event. I also acknowledge informative discussions with A. DiDomenico
on recent results of KLOE. This work is
partially
supported by the European Union through the Marie Curie Research and Training Network \emph{UniverseNet} MRTN-CT-2006-035863.

\end{document}